\documentclass[a4paper]{article}
\usepackage{WCSB,epsfig}
\usepackage{amsmath,amssymb,amsthm}
\def\x{{\mathbf x}}

\def\XX{{\widetilde{X}}}
\def\f{{\mathbf f}}
\def\y{{\mathbf y}}
\def\I{{\mathbb I}}
\def\A{{\mathbb A}}
\def\O{{\mathbb O}}

\def\S{{\mathbb S}}
\newcommand\GF{\ensuremath{\{0,1\}}}
\newcommand{\reff}[1]{(\ref{#1})}

\usepackage{algorithmic}
\usepackage{algorithm}
\usepackage{tikz}
\usepackage{pgf}
\usepackage{pgfplots}
\usepackage{color}
\title{Computing preimages of Boolean Networks}
\name{Johannes Georg Klotz, Martin Bossert, and Steffen Schober}
\address{Institute of Communications Engineering,\\
Ulm University, 
Albert-Einstein-Allee 43,
89081 Ulm, Germany\\
Email: johannes.klotz@uni-ulm.de\\
}
\begin{document}

\maketitle
\begin{abstract}
In this paper we present an algorithm to address the predecessor problem of feed-forward Boolean networks. We propose an probabilistic
algorithm, which solves this problem in linear time
with respect to the number of nodes in the network. Finally, we evaluate our algorithm for random Boolean networks and the regulatory network of \emph{Escherichia coli}.
\end{abstract}
\section{INTRODUCTION}
\label{sec:intro}
In systems and computational biology Boolean networks (BN) are widely used to model regulative dependencies of organisms \cite{KPS03,CKR04}.
We consider networks, which map a set of environmental conditions to the presence of proteins and finally to actual chemical reactions, which are often modeled as fluxes of a \emph{flux-balance analysis} \cite{FHR07}. Hence, these networks are used to make \emph{in silico} predictions of behavior of organisms in a certain environment \cite{FGV12}.


In this paper we address the inverse problem, i.e., we want to predict environmental conditions that allow certain reactions to take place, and others not.
Hence, in general, we need  to find a set of possible inputs that lead to a given output.
This so called \emph{predecessor problem} or \emph{preimage problem} has been addressed by Wuensche in \cite{W94} and has been shown to NP-hard in general \cite{AHZCN09}, which makes it  infeasible to solve it for large networks.
In \cite{KSB12} an algorithm with reduced complexity for BNs with canalizing Boolean functions has been introduced.
However, the problem is infeasible under certain conditions. Both algorithms are designed to find the whole set of preimages, i.e., all inputs to the BN with lead to a certain, desired, output. 

In some applications, knowledge of the whole preimage set is not important, merely it can be sufficient to 
know a subset of the preimage set.
Here, we propose an probabilistic algorithm, which solves this problem in linear time with respect to the number of nodes in the network, 
based on a variation of the well known Sum-Product-Algorithm \cite{KFL01},
which is used for a variety of tasks, including decoding error correction codes in communication engineering~\cite{Gallager1963}.

\section{Boolean Networks and Main Idea}
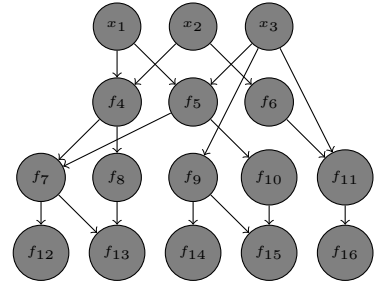
\begin{figure}
\centering
\begin{tikzpicture}[func/.style={draw,circle, font=\tiny, fill=gray}, var/.style={draw,circle, font=\tiny, fill=gray}, flux/.style={draw,circle, font=\tiny, fill=gray}]

\node[func] at (0,0) (f11) {$x_1$}; 
\node[func] at (1,0) (f12) {$x_2$}; 
\node[func] at (2,0) (f13) {$x_3$}; 
\node[func] at (-1,-2) (f21) {$f_7$}; 
\node[func] at (0,-2) (f22) {$f_8$}; 
\node[func] at (1,-2) (f23) {$f_9$};
\node[func] at (2,-2) (f24) {$f_{10}$}; 
\node[func] at (3,-2) (f25) {$f_{11}$}; 

\node[func] at (0,-1) (f01) {$f_4$}; 
\node[func] at (1,-1) (f02) {$f_5$}; 
\node[func] at (2,-1) (f03) {$f_6$};

\node[flux]  at (-1,-3) (y1) {$f_{12}$}; 
\node[flux]  at (0,-3) (y2) {$f_{13}$}; 
\node[flux]  at (1,-3) (y3) {$f_{14}$}; 
\node[flux]  at (2,-3) (y4) {$f_{15}$}; 
\node[flux]  at (3,-3) (y5){$f_{16}$}; 

\draw[->]
(f13) edge (f02)
(f02) edge (f21)
(f02) edge (f24);

\draw[->]
(f11) edge (f01)
(f11) edge (f02) 
(f12) edge (f01) 
(f13) edge (f25)
(f03) edge (f25)
(f13) edge (f23)
(f12) edge (f03)
(f01) edge (f21)
(f01) edge (f22);

\textbf{•}
\draw[->]
(f21) edge (y1)
(f22) edge (y2)
(f23) edge (y3)
(f24) edge (y4)
(f25) edge (y5)
(f23) edge (y4)
(f21) edge (y2);

\end{tikzpicture}
\caption{Example of a Feed-Forward Network}
\label{fig:net}
\end{figure}
We consider networks like shown in Figure \ref{fig:net}, mapping the values of the $N$ in-nodes $\I=\{1,2,3\}$ to the $M$ out-nodes $\O =\{12,13,14,15,16\}$, i.e., we can represent this BN as a function mapping the $N$ input values uniquely to the $M$ output values:
\begin{equation*}
\f: \GF^N \rightarrow \GF^M .
\end{equation*}

The network itself consists of $n$ nodes, and a set of directed edges connecting these nodes. Each node $i$ has a certain state, which can be either zero or one, represented by a variable $x_i$. Its value is determined by evaluating a Boolean function (BF) $f_i$. Further, lets define the set $\tilde{n}(f_j)$ as the incoming nodes of node $j$. For example in Figure~\ref{fig:net}, $\tilde{n}(f_5)=\{1,3\}$. The BF $f_j$ is a function mapping $k_j=|\tilde{n}(f_j)|$ values of $\GF^k$ to $\GF$, where $k$ is also called the in-degree of node $j$. The number of edges emerging from a node is called out-degree. 

Given a vector of input values $\x \in \GF^N, \x = (x_1, x_2, \ldots, x_N)$ the corresponding output of $\f$ is $\y = \f(\x), \y \in \GF^M$. 
In general there does not exist a unique inverse function $\f^{-1}$. Instead the cardinality  of the set $\Omega_y:=\{x:\f(\x)=\y\}$
will be larger one. We call $\Omega_y$ the set of preimages of $y$.

In this paper we are interested to find at least parts of $\Omega_y$. 
Suppose there is a probability distribution $P_\y$ on $\GF^N$ such that 
\begin{equation*}
    P_\y\{ \x \} = \begin{cases}
	\frac{1}{|\Omega_y|}  & ~\text{if}~ \x \in \Omega_y \\
	0 		   & ~\text{else}
    \end{cases}.
\end{equation*}
If we knew the probability distribution  $P_\y$, we would have solved the problem.
But as explained, this is too difficult in general. Our main idea now 
is to approximate $P_\y$ by the product of the marginal distributions $P_i$ on the 
individual $x_i$, i.e.,
\begin{equation*}
    P_\y \approx \prod_{i=1}^N P_{i},
\end{equation*}
as the well-known Sum-Product algorithm can be used to compute the marginals efficiently. 
If the approximation is \emph{good enough} sampling out the product of the marginals will yield an element in $\Omega_y$ with reasonable
probability.

\section{Proposed Algorithm}\label{sec:algo}
In this section we will first discuss the basic principles of factor graphs and the Sum-Product Algorithm (Section \ref{sec:sum_prod}). Then we will describe the BN as factor graph in Section \ref{sec:bp} and will formulate the actual algorithm to find the marginals in Section \ref{sec:input}. Finally, the sampling is described.

\subsection{Factor Graphs and Sum-Product Algorithm} \label{sec:sum_prod}
Assume some function $g(x_1,\ldots, x_n)$ defined on some domain $\A^n$, which can be factorized in $m$ \emph{local} functions $h_j, j \in [m] := \{1,2,\ldots, m\}$, i.e.,
\begin{equation*}
g(x_1, \ldots, x_n) = \prod_j h_j(X_j),
\end{equation*} 
where $X_j$ is the subset of $[n]$ containing the argument of $h_j$. We can then define a factor graph \cite{KFL01} as a bipartite graph consisting of $n$ nodes representing variables $\{x_1,\ldots,x_n\}$ (variable nodes) 
and of $m$ nodes representing functions $\{f_1,\ldots f_m\}$ (function node).  Edges only exist between a function node and a variable node if and only if  $x_i$ is an input to function $f_j$.

The marginal function $g_i(x_i)$  is defined as  {\cite{KFL01}}
\begin{equation*}
g_i(x_i) = \sum_{\sim \{x_i\}} g(x_1,\ldots,x_n),
\end{equation*}
where $\sum_{\sim \{x_i\}} g(x_1,\ldots,x_n)$  is defined as
\begin{align*}
&\sum_{\sim \{x_i\}} g(x_1,\ldots,x_n) \\&= \sum_{x_1 \in \A} \ldots\sum_{x_{i-1} \in \A} \sum_{x_{i+1} \in \A} \ldots \sum_{x_n \in \A} g(x_1,\ldots,x_n),
\end{align*}
In general the computation of the $g_i$ is difficult, but due to the factorization of $g$ the task can be efficiently solved using the 
the so called Sum-Product algorithm \cite{KFL01}. 
The algorithm iteratively passes \emph{messages} between the nodes of the graph.
At each iteration the messages $\mu$ are sent from the function nodes to the variable nodes, containing the corresponding marginal function of the local function. These messages are computed as follows \cite{KFL01}:

\textbf{function to variable node:}\\
\begin{equation*}
\mu_{h\rightarrow x}(x) = \sum_{\sim\{x\} }\left( h(n(h)) \prod_{y\in n(h)\setminus \{x\}} \lambda_{y\rightarrow h}(y)\right),
\end{equation*}
where $n(i)$ give the set of neighboring nodes of node $i$.

At the variable nodes, these messages are then combined to a marginal function $\lambda$ and sent back to the function nodes \cite{KFL01}:

\textbf{variable to function node:}\\
\begin{equation*}
\lambda_{x\rightarrow h}(x) = \prod_{ q \in n(x)\setminus \{h\}} \mu_{q\rightarrow x}(x).
\end{equation*}

\subsection{The Boolean Network as Factor Graph.}\label{sec:bp}
We apply the concept of factor graphs to BNs. Each node in the network represents one variable $x_i \in \GF, i \in [n]$ of the factor graph, hence we have $n$ variable nodes. Each BF $f_j$ of the BN ($j\in  [n] \setminus \I$) is a function node and is connected to the node $j$ and the incoming nodes $\tilde{n}(f_j)$. Lets to define $\XX_j$ as the variables of the incoming nodes of node $j$, i.e. the argument of the BN $f_j$. Further, we define $\XX_j^{(i)}$ as $\XX_j$ without the node $i$.

Finally, if we consider the variables as each node as random variables, we have a common distribution of all variables nodes described by the density function,
$$
g_{x_1,\ldots,x_n}(x_1,\ldots,x_n) \equiv g(x_1,\ldots,x_n),
$$
For sake of readability we will omit the subscript of the density function, if they are obvious from context.
We are interested in finding the marginal distributions of the in-nodes, which can be described by the
density functions
\begin{equation*}
g_{x_i}(x_i)= \sum_{\sim x_i}g_{x_1,\ldots,x_n}(x_1,\ldots,x_n) \text{~~~} \forall i \in \I.
\end{equation*}
This problem is an instance of the problem described in 
in Section~\ref{sec:sum_prod}, hence, we apply the same methods here.

\subsubsection{Update Rule: function to variable node}
If we focus on one function node $j\in [n]\setminus \I$ there exists a common distribution of all variables relevant for this node. Namely, these relevant variables are the ones located in $\XX_j$ of the BF $f_j$, and the value of node $j$. We can write the density of this distribution as:
$$
p(x_j,\XX_j).
$$
Lets define $\tilde{n}(f_j)$ as the set of indices of the input nodes of the BF $f_j$.

We need to send the local marginal distribution of each variable $i \in \{j\} \cup \tilde{n}(f_j)$ back to the variable node, or more formal:
\begin{equation} \label{eq:fu-va:gen}
\mu_{j\rightarrow i}(x_i) = \sum_{\sim \{x_i\}} p(x_j,\XX_j)= \sum_{\sim \{x_i\}} p(x_j,x_i,\XX_j^{(i)})
\end{equation}
If $i=j$ , i.e. if the message is designated for the node containing the output of the BF, the density of the marginal distribution becomes:
\begin{align*}
\mu_{j\rightarrow j}(x_j) &= \sum_{\sim \{x_j\}} p(x_j|\XX_j) \cdot p(\XX_j) \nonumber \\ 
	&=\sum_{\sim \{x_j\}} f_j(\XX_j)\cdot p(\XX_j)
\end{align*}
which is the probability distribution of the functions output.
We can assume that the elements of $\tilde{X}_j$ are pairwise independent, hence, we can write:
$$
p(\XX_j) = \prod_{l\in \tilde{n}(f_j)} \lambda_{l}(x_l),
$$
where $\lambda_{l}$ is the probability distribution of node $j$ and is defined in Section~\ref{sec:va-fu}.

In the other cases, i.e., $i\neq j$, Eq.~\reff{eq:fu-va:gen} becomes:
\begin{equation*}
\mu_{j\rightarrow i}(x_i) = \sum_{\sim \{x_i\}} p(x_i|x_j,\XX_j^{(i)}) \cdot p(x_j,\XX_j^{(i)}).
\end{equation*}
We still can assume that the elements of $\XX_j^{(i)}$ are pairwise independent, hence, we can write:
\begin{align*}
p(x_j,n(f_j)\setminus x_i) &= p(x_j|\XX_j^{(i)}) \cdot p(\XX_j^{(i)}) \\
&=p(x_j|\XX_j^{(i)}) \prod_{l\in \tilde{n}(f_j) \setminus \{j\}} \lambda_{l}(x_l).
\end{align*}

If the Boolean functions output $x_j=f_j(\XX_j)$ is already completely determined by $\XX_j^{(i)}$, i.e., if the variable $x_i$ has no influence on the output for this particular choice of the other variables, we assume $x_i$ to be uniformly distributed:
$$
p(x_i|x_j,\XX_j^{(i)}) = \frac{1}{2}
$$
and since $x_j$ is completely determined by $\XX_j^{(i)}$
$$
p(x_j,\XX_j^{(i)}) = \prod_{l\in \tilde{n}(f_j) \setminus \{j\}} \lambda_{l}(x_l).
$$
Otherwise, $x_i$ is totally determined by $x_j$ and the other variables, i.e., $x_i$ is $0$ or $1$ depending on BF. Hence, we can write 
\begin{equation*}
p(x_i|x_j,n(f_j)\setminus x_i) =  p_{x_j}(f(\XX_j^{(i)},x_i)=x_j),
\end{equation*}
where $p_{x_j}(f(\XX_j^{(i)},x_i)=x_j)$ is either $0$ or $1$.
Further we can assume $x_j$ independent of $\XX_j^{(i)}$, hence,
\begin{align*}
p(x_j,\XX_j^{(i)})= \lambda_{j}(x_j) \prod_{l\in \tilde{n}(f_j) \setminus \{j\}} \lambda_{l}(x_l).
\end{align*}
Finally, we can summarize for $i\neq j$:
\begin{equation} \label{eq:fu-va:neq}
\mu_{j\rightarrow i}(x_i) = \sum_{\sim \{x_i\}} \xi_{i,j} \prod_{l\in \tilde{n}(f_j) \setminus \{j\}} \lambda_{l}(x_l),
\end{equation}
with 
$$
\xi_{i,j} =
\begin{cases}
\frac{1}{2} &\text{, if }f_j(\XX_j^{(i)},x_i=0)=f_j(\XX_j^{(i)},x_i=1) \\
\lambda_j(x_j)&\text{, else} \\
\end{cases}.
$$
\subsubsection{Update Rule: variable to function node}\label{sec:va-fu}
The update rule is the same for all variable nodes $j \in [n]$ and is independent of the function node to which they are directed. 
\begin{equation} \label{eq:va-fu}
\lambda_{j}(x_j) = \prod_{l\in\S_j}\mu_{l\rightarrow j}(x_j),
\end{equation}
where $\S_j$ is the set of all function nodes, which have node~$j$ as input.
\subsection{Finding the Input Distributions}\label{sec:input}
In our algorithm, we use the well known log-likelihood ratio (LLR)  to represent the probability distribution of binary variables \cite{Hagenauer1996}. It is defined as:
\begin{equation}\label{eq:llr}
L_X=\ln\frac{p(x=0)}{p(x=1)}.
\end{equation}
A scheme of the algorithm is given in Algorithm \ref{alg:1}. 

The probability distribution of each node $j \in [n]$ at iteration $t$ is given as $L_j^{(t)}$ and are initialized with $L_j^{(0)}=0$, which is equivalent to the uniform distribution. Then we set the LLRs for the out-nodes to either $-\infty$ or $+\infty$ depending on the desired output $\y$ of the BN.
At each iteration the algorithm can be split in two steps. The first step iterates over all function nodes $j \in [n] \setminus \I$ and all input variables $i\in \tilde{n}(f_j)$ calculating the LLR $L_{j\rightarrow i}^{(t)}$ using Eq. \reff{eq:fu-va:neq} and Eq. \reff{eq:llr}. 

In the second step we update all variables-nodes, where the LLRs $L_j$ represents the distributions $\lambda_j$ and, hence, the product of Eq. \ref{eq:va-fu} becomes a summation. Please note, that the LLR of the previous iteration is also added to the sum, in order to prevent rapid changes of the distributions.

After performing a certain number of iterations $t_{max}$, the desired marginal distributions of the input variables are found.

\begin{algorithm}
\caption{ }\label{alg:1}
\begin{algorithmic}
\STATE Initialize $L_j^{(0)}=0$ for all nodes
\STATE Set the desired LLRs of the out-nodes, i.e., $L_j^{(0)}$ is either $-\infty$ or $+\infty$, for all out-nodes $j\in \O$.
\STATE t=0
\REPEAT 
\STATE t=t+1
\FOR {each non-in-node $j\in [n]\setminus \I$}
\FOR {each input variable $i \in \tilde{n}(f_j)$} 
\STATE calculate $L_{j\rightarrow i}$ using Eq. \reff{eq:fu-va:neq} and Eq. \reff{eq:llr}
\ENDFOR
\ENDFOR
\FOR {each non-out-node $v$} 
\STATE $L_{j}^{(t)}=L_{j}^{(t-1)}+\sum_{l \in \S_j} L_{l\rightarrow j}^{(t)}$
\ENDFOR
\UNTIL maximum number of iterations reached
\end{algorithmic}
\end{algorithm}
\subsection{Sampling}\label{sec:sampling}
The sampling part of our approach is straight forward. Using the marginal distributions $L_j^{(t_{max})}, j\in \I$ we randomly draw vectors $\x$ and check if they fulfill $\y=f(\x)$. If so, they are added to the set $\widetilde{\Omega}_\y$. This procedure is repeated for a certain number of samples.
\section{SIMULATION RESULTS AND DISCUSSION}\label{sec:simo}
We tested our algorithm with randomly generated networks and the regulatory network of \emph{Escherichia coli} (\emph{E-coli})  \cite{CKR04}.
The random networks consist of $2400$ nodes with $N=200$ and $M=1200$. 
We have chosen the BFs from:
\begin{itemize}
\item all functions with $k\leq 15$ (Type A)
\item unate, i.e. locally monotone, functions with $k\leq 15$ (Type B)
\end{itemize}
After generating a network we draw a certain number $T$ of uniformly distributed input vectors $\x$ and obtain $\y=\f(\x)$. For each $\y$ we applied then Algorithm~\ref{alg:1} to obtain the marginal distributions $L_j^{(t_{max})}, j \in \I$.

To investigate the convergence behavior with respect to  $t_{max}$ and we first apply hard-decision to evaluate a good choice for $t_{max}$, i.e., we generate an estimate $\tilde{\x}$ by setting 
$$
\tilde{x}_j = 
\begin{cases}
0 &\text{~if~} L_j^{(t_{max})} >0 \\
1 &\text{~if~} L_j^{(t_{max})} <0
\end{cases}
$$

\begin{figure}
  \begin{tikzpicture}

    \begin{axis}[
    xlabel=$t_{max}$, 
    ylabel=$Similarity$, 
    xmin=0.0, 
    legend pos=south east, 
    legend style={cells={anchor=east}}, 
    width=212pt, 
    height=7cm, 
    enlarge x limits=false, 
    enlarge y limits=false, 
    grid=major 
    ]
    \addplot[color = black, mark=+] file {./typeat.dat};
    \addlegendentry{Type A}
    \addplot[color = black, mark=*] file {./typebt.dat};

    \addlegendentry{Type B}
    \addplot[color = black, mark=o] file {./ecolit.dat};

    \addlegendentry{\emph{E-coli}}
    \end{axis}
  \end{tikzpicture}

\caption{Similarity of $\y$ and $\tilde{\y}$ verses $t_{max}$}\label{fig:findt}
\end{figure}
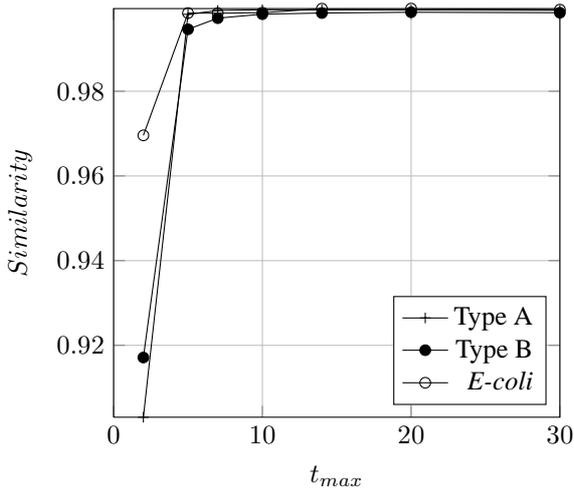

Then we evaluate the network $\tilde{\y}=\f(\tilde{\x})$, and measure the similarity between $\y$ and $\tilde{\y}$ by counting the equal entries and divide them by the length of $\y$. We did so for 100 networks of Type A and B, and set $T=100$. The averaged results can be seen in Figure \ref{fig:findt}.
One can see, that for $t_{max} \geq 14$ there is almost no improvement in the similarity. This number is equal to two times the number of nodes between input and output, i.e., it seems to be sufficient that the messages travel once through the network and back. Thus, the following simulations have been perform setting $t_{max}=14$.

Next, we apply sampling as described in Section \ref{sec:sampling}. We did so for 100 different networks of Type A and B, and the \emph{E-coli} network. For each random network we did $T=100$ runs, for \emph{E-coli} $T=1000$. The results can be viewed in Table \ref{tab:1}. We depict the percentage of solved networks, i.e. the portion of networks we found at least one valid $\x \in \Omega_\y$. Further, we give the average number of valid $\x$ and the average number of unique $\x$.
\begin{table}[h!]

\begin{tabular}{l|c|c|c|c}
network & num of samples & solved & valid & unique\\
Type A &1000 & 89\%& 608.81 & 4.43  \\
Type B &1000 & 95.9\%  & 270.74	& 68.60 \\
\emph{E-coli} &1000& 98.6\% & 193.3 & 193.3
\end{tabular}
\caption{Simulation results for different networks}\label{tab:1}
\end{table}

One can see from the results, that in general for most networks and $\y$s at least one preimage can be found. It is worth mentioning, that for the \emph{E-coli} network every sampled solution was unique. This is due to the fact, that there exist a few inputs, who completely determine the output. The other input variables have then no influence and hence a marginal distribution of $0.5$. Further, the results for the network of type B are much better than for type A. It seems that the marginal distributions for unate functions give better estimation of the actual distribution than the marginal distributions for non-unate functions.
\section{ACKNOWLEDGMENTS}
The authors would like to thank Shrief Rizkalla for implementing parts of the simulation. This work was supported by the German research council "Deutsche Forschungsgemeinschaft" (DFG) under Grant Bo 867/25-2.

\bibliographystyle{IEEEbib}
\bibliography{literatur}

\end{document}